\def \SAIT #1 #2 {{\em Mem.\ Soc.\ Astron.\ It.\/} {\bf #1}, #2}
\def \MESS #1 #2 {{\em The Messenger\/} {#1}, #2}
\def \ASTRNACH #1 #2 {{ Astron. Nach.\/} { #1}, #2}
\def \AAP #1 #2 {{ A{\rm \&}A\/} {#1}, #2}
\def \AAL #1 #2 {{ A{\rm \&}A\/} {#1}, L#2}
\def \AAR #1 #2 {{ A{\rm \&}AR\/} {#1}, #2}
\def \AAS #1 #2 {{ A{\rm \&}AS\/} {#1}, #2}
\def \AJ #1 #2 {{ AJ\/} {#1}, #2}
\def \ANNREV #1 #2 {{ ARA{\rm \&}A\/} {#1},#2}
\def \APJ #1 #2 {{ ApJ\/} {#1}, #2}
\def \APJL #1 #2 {{ ApJ\/} {#1}, L#2}
\def \APJS #1 #2 {{ ApJS\/} {#1}, #2}
\def \APSS #1 #2 {{ Ap{\rm \&}SS\/} {#1}, #2}
\def \ASR #1 #2 {{ Adv. Space Res.\/} {#1}, #2}
\def \BAIC #1 #2 {{ Bull. Astron. Inst. Czechosl.\/} { #1}, #2}
\def \JSQRT #1 #2 {{ J. Quant. Spectrosc. Radiat. Transfer\/} {
#1}, #2}
\def \MN #1 #2 {{ MNRAS\/} { #1}, #2}
\def \MEM #1 #2 {{ Mem. R. Astr. Soc.\/} { #1}, #2}
\def \PLR #1 #2 {{ Phys. Lett. Rev.\/} { #1}, #2}
\def \PASJ #1 #2 {{ Publ. Astron. Soc. Japan\/} { #1}, #2}
\def \PASP #1 #2 {{ Publ. Astr. Soc. Pacific\/} { #1}, #2}
\def \NAT #1 #2 {{ Nat\/} { #1}, #2}
\def \ACTA #1 #2 {{ Acta Astron.\/} { #1}, #2}
\documentclass[preprint]{emulateapj}

\DeclareRobustCommand{\ion}[2]{%
\relax\ifmmode
\ifx\testbx\f@series
{\mathbf{#1\,\mathsc{#2}}}\else
{\mathrm{#1\,\mathsc{#2}}}\fi
\else\textup{#1\,{\mdseries\textsc{#2}}}%
\fi}

\slugcomment{To appear in ApJ.}

\shorttitle{The role of the synchrotron component in the MIR spectrum of M~87}
\shortauthors{L. Buson et al.}

\usepackage{psfig,latexsym,longtable,lscape,rotating}

\def\smallskip{\vskip 6pt}

\def\M12{${\rm M_{12}}$}

\begin{document}

\title{The role of the synchrotron component in the mid infrared spectrum of M~87}
\author{L. Buson\altaffilmark{1},
A. Bressan\altaffilmark{1,2,3},
P. Panuzzo\altaffilmark{4},
R. Rampazzo\altaffilmark{1},
J. R. Vald\'es\altaffilmark{3},
M. Clemens\altaffilmark{1},
A. Marino\altaffilmark{5},
M. Chavez\altaffilmark{3},\\
G. L. Granato\altaffilmark{6},
L. Silva\altaffilmark{6}
}
\altaffiltext{1}{INAF Osservatorio Astronomico di Padova, Vicolo
dell'Osservatorio 5, 35122 Padova, Italy; lucio.buson@oapd.inaf.it}

\altaffiltext{2}{Scuola Internazionale Superiore di Studi Avanzati (SISSA),
via Beirut 4, 34014, Trieste, Italy}

\altaffiltext{3}{Instituto Nacional de Astrof\'{\i}sica, Optica y Electr\'onica,
Apdos. Postales 51 y 216, C.P. 72000 Puebla, Pue., M\'exico}

\altaffiltext{4}{CEA/DAPNIA/Service d'Astrophysique CEA Saclay
91191 Gif sur Yvette Cedex, France}
\altaffiltext{5}{Johns Hopkins University, Dept. of Physics and Astronomy
3400 N. Charles St.  Baltimore, MD 21218 U.S.A.}

\altaffiltext{6}{INAF Osservatorio Astronomico di Trieste, Via Tiepolo 11, I-34131 Trieste, Italy}

\date{Received / Accepted }

\begin{abstract}
We study in detail the mid-infrared {\it Spitzer}-IRS spectrum of M~87
in the range 5 to 20 $\mu$m.
Thanks to the high sensitivity of our {\it Spitzer}-IRS spectra
we can disentangle the stellar and nuclear components of this active galaxy.
To this end we have properly subtracted from the M~87 spectrum,
the contribution of the  underlying stellar continuum,
derived from passive Virgo galaxies in our sample.
The residual is a clear power-law, without any
additional thermal component,
with a zero point consistent with that obtained by
high spatial resolution, ground based observations. The residual is independent
of the adopted passive template.
This indicates that the 10$\mu$m silicate emission shown in spectra of
M~87 can be entirely accounted for by the underlying
old stellar population, leaving little room for a possible
torus contribution.
The MIR power-law has a slope $\alpha\sim$ 0.77-0.82 (S$_\nu\propto\nu^{-\alpha}$),
consistent with optically thin synchrotron emission.

\end{abstract}


\section{Introduction}

\citet{Bressan06} presented high signal-to-noise ratio
 {\it Spitzer} Infrared Spectrograph (IRS)
observations of 18 Virgo early-type galaxies (ETGs hereafter), selected from those that
define the color-magnitude relation of the cluster, with the aim of detecting the
silicate emission of their dusty, mass-losing evolved stars.
They found that 13 ETGs, i.e.76\% of the sample, show
a pronounced broad silicate feature that is spatially extended and likely of stellar
origin, in agreement with model predictions of old populations \citep{Bressan98}.
The absence of any other feature in their spectra suggested that these galaxies
could be considered as the prototype of passively evolving ETGs.
In contrast, the IRS spectrum of the other four ETGs presented various levels of activity.

Among these, NGC~4486 (M~87), the well-known giant elliptical
located at the cluster centre, stands out. Its mid-infrared (MIR hereafter)
 spectrum, besides narrow atomic emission lines,
shows an excess of emission at longer wavelengths with respect to passive ETGs.
The silicate emission at 10$\mu$m is also evident in the spectrum, and since
M~87 hosts a strong central radio source (Virgo~A) and a jet, visible from radio
to X-ray wavelengths, it could be an indication of possible torus emission,
 as already found in other AGNs (Siebenmorgen et al. 2005, Shi et al. 2006,
 Perlman et al. 2007).

The importance of the torus in the context of the AGN energetics makes the
debate about the origin of the nuclear MIR emission in M~87 very intriguing.
Early, high spatial resolution, ground-based MIR observations
of M~87 \citep{Perlman01,Whysong04} have shown that
the emission from the nucleus is consistent with synchrotron emission.
Both investigations have excluded the possibility that M~87 hosts a quasar-like nucleus
similar to that at the center of Cen~A. However, neither could exclude the possibility
of a low-luminosity AGN. Indeed \citet{Perlman01}
detected a faint, extended, circum-nuclear component that could originate in a torus
contributing $\sim$7\% of the nuclear luminosity.
Alternatively, they suggested
that this extended component could be stellar emission from red giant stars.
More recently, \citet{Perlman07} reanalyzed this issue by means of
high resolution ground based imaging and {\it Spitzer} spectroscopy.
They confirmed the consistency with synchrotron emission for the MIR fluxes of the
knot A/B complex in the jet. However, they
measured a spectral index $\alpha_{IR}\sim$0.41, much lower than before
and different from that found in the nearby structures \citep[e.g.][]{Shi07}.
They still noticed deviations from the pure synchrotron emission
and speculated that, at least in the shorter wavelength range, this could
 be due either to contribution from M and K stars or contamination from the
low frequency tail of the emission from knot HST-1,
not resolved by {\it Spitzer} but also not apparent in the 11.7$\mu$m
{\it SUBARU} image of the nucleus.

We revisit here the nature of the MIR emission of M~87
exploiting our high signal-to-noise {\it Spitzer} spectra of passively evolving ETGs.
Taking advantage of these data we have constructed very accurate MIR
templates of the stellar populations in these galaxies (Valdes et al. 2009, in prep.).

\emph{These templates will be used to disentangle the nuclear
non-stellar emission of M~87 by a proper subtraction process.}

In section \ref{data} we summarize our {\it Spitzer IRS} observations and data extraction as well as
all supplementary data needed to perform the subtraction.
In section \ref{subtraction} we describe the subtraction of the spectral
energy distribution (SED) of a passive ETG from the IRS spectrum of M~87 while, in  section \ref{residual}
we discuss the properties of the residual component.
Our conclusions are presented in section \ref{conclusions}.
\begin{figure}
\center{\resizebox{0.40\textwidth}{!}{
\psfig{figure=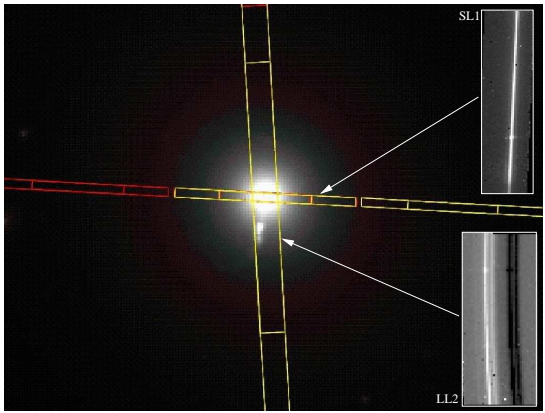,clip=}
}}
\caption{IRS SL (nearly horizontal) and LL (nearly vertical)
slits superimposed on the MIPS image of M 87 at $\lambda$=24$\mu$m.
The two additional panels on the right side show a portion of the LL2 (bottom) and SL1 (top)
2-D spectral images. Notice that the former does clearly show both the spectra of the nucleus and that of
the nearest bright knots,
whereas the SL slit is not affected by these components. The extraction width of the LL spectra
does not include the contribution from these knots.}
\label{fig1}
\end{figure}
%
\section{Observations}
\label{data}
In this section we describe how we have obtained the NIR/MIR SEDs of
M~87, and of  three other elliptical galaxies, NGC~4365, NGC~4382 and NGC~4636.
These were selected to be "passive", insofar as they show no evidence of either recent past star formation or AGN activity.
They will be used as templates for the stellar component in our attempt to
isolate the non stellar emission of the nucleus of M~87.

\subsection{Spitzer IRS}
Standard staring mode short (SL1 and SL2) and long (LL2) low resolution
IRS spectral observations of M~87 and other selected Virgo cluster ETGs
were obtained during the
first {\it Spitzer} General Observer Cycle on 2005 June 1 as part of
 program ID~3419 (PI A. Bressan).
The spectra were extracted in a fixed aperture (3.6"$\times$18" for SL and
10.2"$\times$10.4" for LL), using a purpose built program that
allows the reconstruction of the intrinsic galaxy profile \citep{Bressan06}.

Position angles of the
SL (3.6" wide) and LL (10.2" wide) IRS slits in the case of M~87,
are shown
superimposed to the MIPS 24$\mu$m image, in Figure~\ref{fig1}.
At these wavelengths the emission of NGC~4486 appears quite complex
even at the low resolution of MIPS  24$\mu$m.
Perlman et al. (2001) have provided a thorough description of
the MIR  spatial emission of the
central regions of M87.
Their 10.8~$\mu$m image shows
the ``nucleus'' and the D, F and A, B and C knots, whose
fluxes are reported in their Table 1.
To check which of these external components can affect our spectra
we show the 2-D spectral images of
of the LL2  and SL1 segments in
the bottom and top panels of Figure \ref{fig1}, respectively.
In the LL2 slit  we can easily recognize both the spectra of the nucleus and that of
the nearest bright A, B and C knots,
whereas the SL1 slit is clearly not affected by these three components.
Thus, the only  extra-nuclear component that may contribute
to our SL spectrum is knot D. However this feature is located 3" from the
nucleus, has low level flux (7\% of the nucleus, Perlman et al. 2001)
and thus is likely to provide only a negligible contribution to the SL spectrum.
The other components, located at greater distances,
are well outside the SL slit.
On the other hand our LL slit contains the strong knots
A, B and C that are clearly visible in the spectral images.
In this case the extraction is limited to a width of 10.4"
in order to avoid these strong knots that are not present in
SL. As we will see later our results are independent of the inclusion or
exclusion of the LL segment from our analysis.

The estimate of the flux {\sl emitted} by an extended source within a fixed
slit aperture  involves the deconvolution of the received
flux with the PSF of the instrument. This correction is important
to obtain the shape of the intrinsic spectral energy distribution (SED)
of the galaxy, because from the slit loss correction we have estimated that
for a point source the losses due to the PSF amount to
$\sim$20\% at 5$\mu$m and to $\sim$40\% at 15$\mu$m. Conversely,
a uniform source will suffer no net losses.
In order to recover the intrinsic SED we have convolved
a surface brightness profile model with the instrumental
PSF, and we have simulated the corresponding linear profile
along the slits, taking into account the relative
position angles of the slits and the galaxy.
The adopted intrinsic model profile is a wavelength dependent two dimensional
modified King law (Elson at al. 1987):
\begin{equation}
I \equiv I_0 / \left[1+\frac{X^2}{R_{\rm C}^2}+\frac{Y^2}{(R_{\rm C}\times b/a )^2}\right]^{-\gamma/2}
\label{elson}
\end{equation}
 where $X$ and $Y$ are the coordinates along the major and minor axis
of the galaxies,
$b/a$ is the axial ratio taken from the literature.
$I_0$, $R_{\rm C}$ and $\gamma$
are free parameters that are
functions of the wavelength and are obtained by fitting the
observations with the simulated profile.
In order to get
an accurate determination of the parameters of the profiles several
wavelength bins have been co-added.
{\bf This procedure allows us (i)
to recognize whether a particular feature is resolved
or not, (ii) to reconstruct the intrinsic profile of the source
and (iii) to account for the wavelength dependent PSF losses,
in a more general way than for point like sources.}

\begin{figure*}
\center{
\includegraphics[width=0.35\textwidth,angle=90]{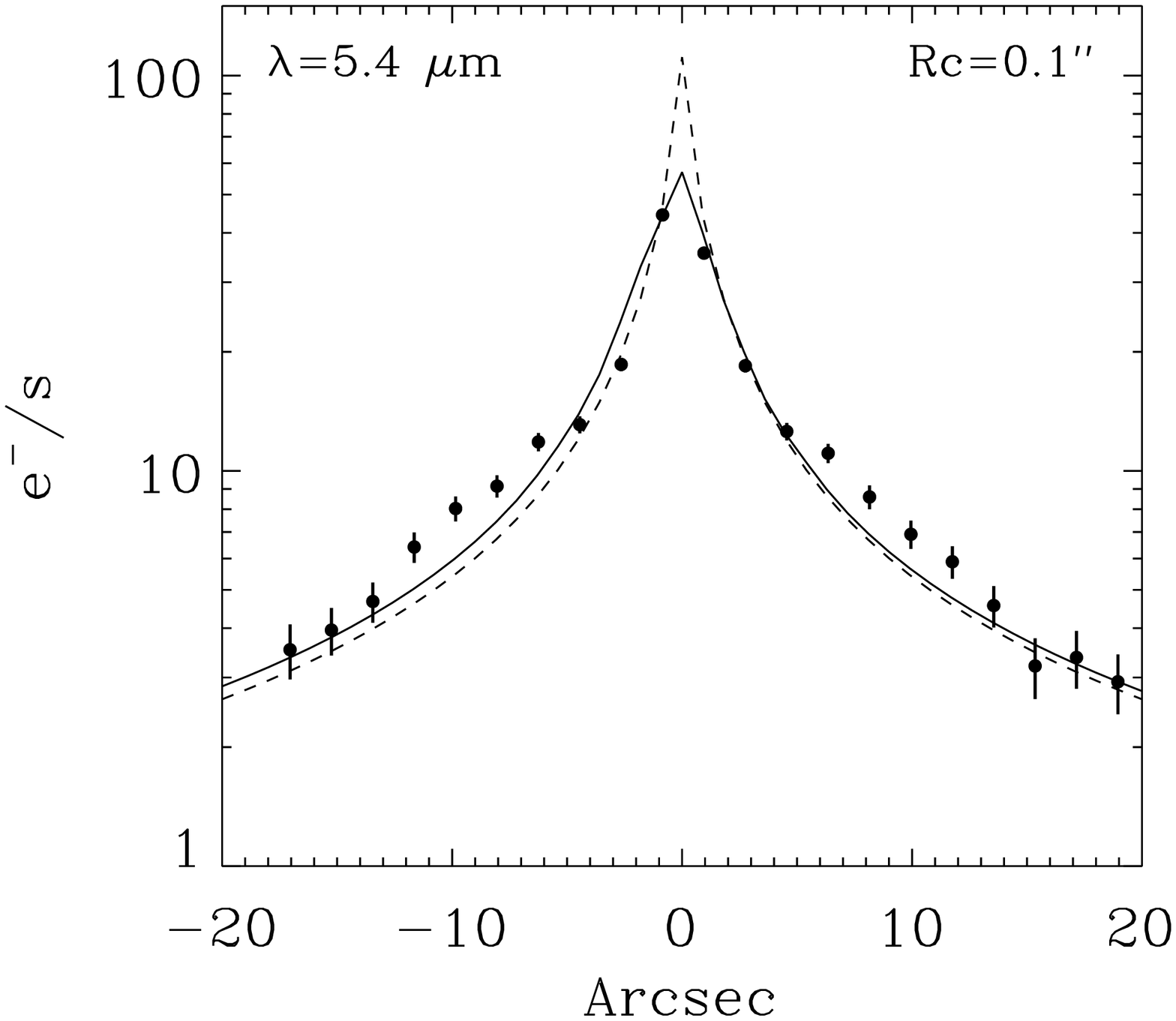}
\includegraphics[width=0.35\textwidth,angle=90]{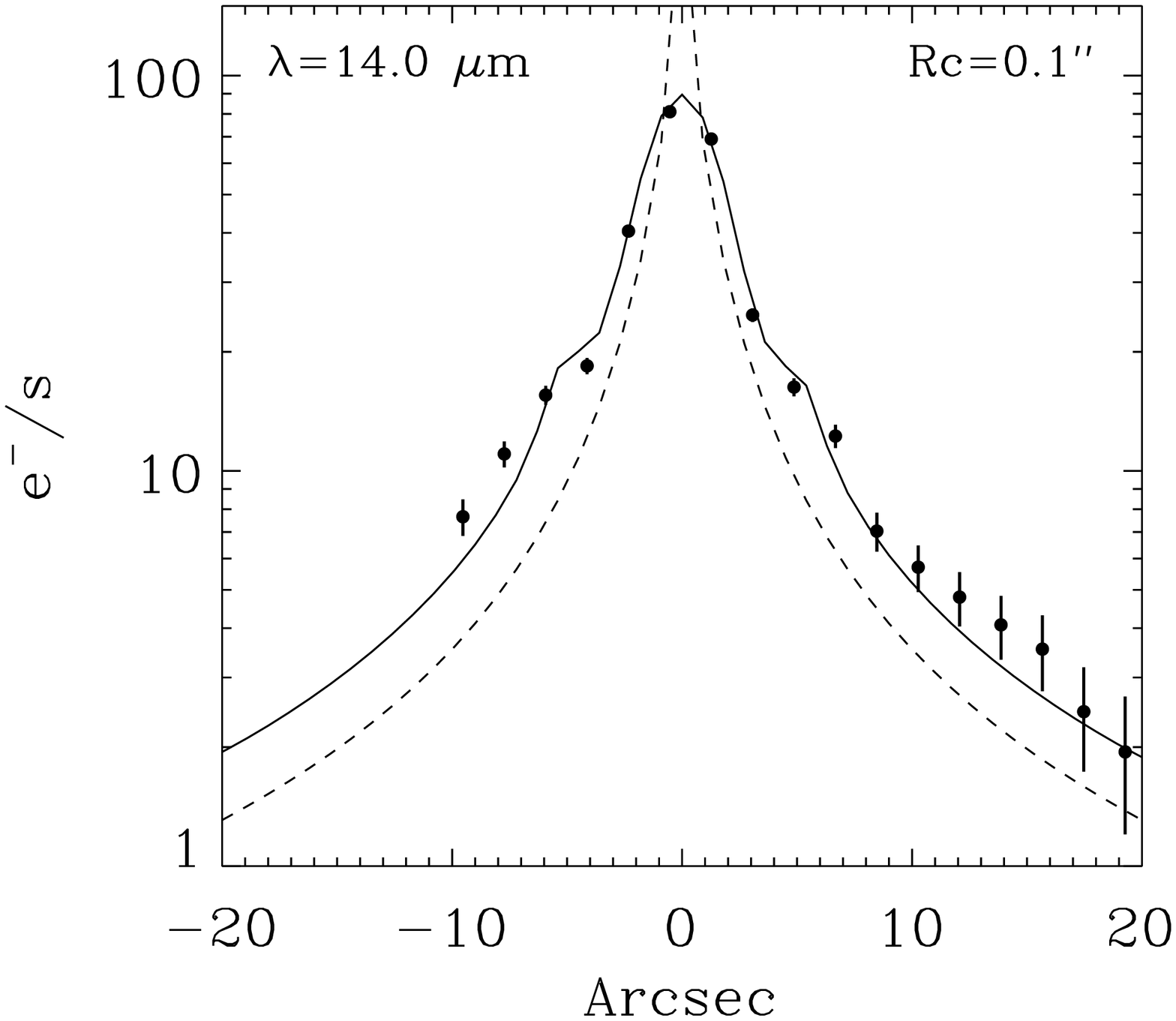}
}
\caption{NGC 4486 spatial emission profiles (filled dots), in electrons/s,
at two selected wavelengths.
The dashed lines show the reconstructed
one dimensional intrinsic spatial emission profiles
sampled along the slit.
The solid lines represent the corresponding one dimensional PSF convolved models,
that are fitted to the data.
Error bars are 1-$\sigma$ uncertainties.}
\label{fig2}
\end{figure*}

Examples of the fitting procedure for NGC 4486
at two selected wavelengths are  shown in Fig.\ref{fig2}.
The dashed lines show the reconstructed
one dimensional intrinsic spatial emission profiles (in electrons/s)
sampled along the slit, while the solid lines represent the corresponding one dimensional PSF convolved models
that are fitted to the data.
The error bars in the figure show 1-$\sigma$ uncertainties that were computed
taking into account the Poissonian noise from the source and the background,
the readout noise and dark current noise, as described in the
Spitzer Observer Manual.
The spatial emission parameters obtained for all the galaxies used in this paper are shown in Table~1.
At short  IRS wavelengths the spatial emission of NGC~4486 appears complex.
It is a superposition of an unresolved central
component and an extended component.
Indeed, in fitting only a single component to the data we find an unresolved central source with
$R_{\rm C}$=0.1". Nonetheless, there is an evident excess over this unresolved component beyond
 $\sim 4^{\prime\prime}$, that does affect our rectangular extraction, 18" long.
The excess remains visible up to $\sim$14$\mu$m beyond which the profile is clearly that of the PSF
($R_{\rm C}$=0.1"), and the excess in the wings is much less important.
From this spatial analysis we conclude that
NGC 4486 is progressively more dominated by un unresolved nuclear component
as the wavelength increases. However, the presence of an underlying
extended emission, evident at 5 $\mu$m,
can be detected even at 9$\mu$m.
Since we have shown that the external jet does not contribute
to the SL segment,
we argue that this emission is of stellar origin.







\subsection{Other Observations}
%
\label{subtraction}
\begin{deluxetable}{lrrrrr }
\tablecaption{MIR Spatial emission parameters of selected early type galaxies}
\label{tab0}
\tabletypesize{\scriptsize}
\startdata
\hline
 &&5.4$\mu$m&&14.0$\mu$m\\
GALAXY       & $R_{\rm C}$&$\gamma$& $R_{\rm C}$&$\gamma$\\
\hline
NGC 4486  &0.1&0.5&0.1&0.7\\
NGC 4365  &3.5&2.0&4.0&2.2\\
NGC 4382  &1.3&1.4&1.6&1.3\\
NGC 4636  &3.7&0.8&3.5&0.9\\
\enddata
\end{deluxetable}
IRAC fluxes were measured from archival ``post basic calibrated data''
images using a customized program written in {\sc IDL}. Before
flux measurements, the background in each image was determined as the
modal sky value of the off-source portion of each image. Fluxes were
then measured in circular apertures centred on each galaxy.
The error on each flux measurement was calculated as the quadrature
sum of the calibration error (taken to be 10\%) and the standard
deviation of the background value. The calibration error is the
dominant term.

In order to place M~87 IRAC and IRS data on the same flux scale, we measured
IRAC 5.8~$\mu$m fluxes in the circular aperture
containing the same flux sampled by IRS at the same wavelengths
This aperture has a radius of 4\arcsec\ and has been selected as the reference aperture
for any band.
Consequently, all broad band fluxes in all other galaxies
were  measured within the same aperture
and their IRS spectra were normalized to
their flux in IRAC channel 3.

In the case of M~87, we have also measured the emission at 24$\mu$m
from the MIPS image, which
may serve as a useful check for our IRS extraction procedure.
Since the IRS extraction procedure indicates that
toward these wavelengths M87 is dominated by the nuclear component, we have taken into account
the effects of the PSF to obtain the MIPS 24$\mu$m flux.
To this end we have
divided the flux measured within the nuclear $4^{\prime\prime}$ radius aperture
with the fractional
energy of the MIPS 24$\mu$m PSF image, encircled by the same radius.

NIR images of the sample galaxies in the J, H and K$_s$ bands
were downloaded from the All-Sky Data Release from the Two Micron All Sky Survey.
\footnote {NASA/IPAC Infrared Science Archive for NASA Infrared and Submillimeter Data
chartered to serve calibrated science products from NASA's infrared and submillimeter
missions. 2MASS Image Services provide the users calibrated images that contain
photometric zero point information in their headers and are suitable for quantitative
photometric measurements.}

Using these data, in addition to our IRS spectra, we have obtained the NIR/MIR SED of
M87 and of the other three ETGs that will be used as templates for the subtraction of
the stellar component.

\section{Stellar component subtraction}
\label{subtraction}
In order to disentangle the nuclear and the extended component
we resort to the comparison of the high signal-to-noise
spectral energy distribution (SED) of M~87
with that of the passive ellipticals we have also observed with Spitzer.
Here we describe the process of subtraction of the stellar component.

\begin{figure}
\centering
\includegraphics[width=0.35\textwidth,angle=90]{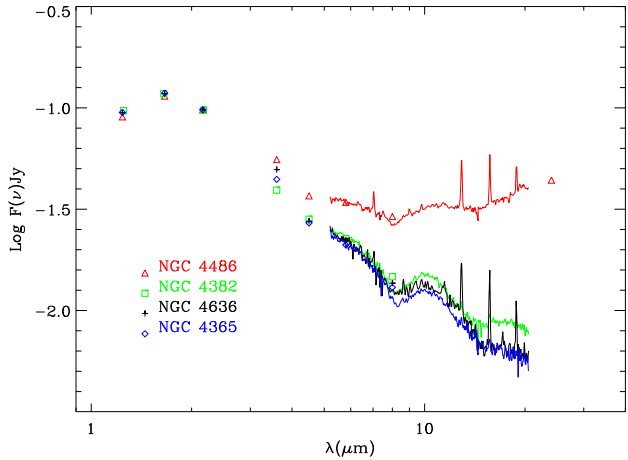}
\caption{Near-IR/MIR SED of M~87 (NGC~4486) and of the other three passive galaxies
 in a circular aperture of 4\arcsec\ radius.
 The passive ETGs have been normalized
to the Ks broad band flux of M~87.}
\label{fig3}
\end{figure}

Before subtraction, all the template galaxy SEDs were normalized to the flux of M~87 in the 2MASS Ks band.
This band was selected because it is the nearest band to the IRS
spectrum with a negligible non-stellar component.
The NIR/MIR  SED of M~87 and of the template ETGs, after this normalization, are shown in Figure \ref{fig3}.

We notice that all SEDs are very similar in the NIR bands.
This similarity, that extends to optical wavelengths,
indicates a uniformity of the underlying stellar populations. The SEDs of passive
ETGs are very similar also in the MIR, especially where
the integrated spectrum is dominated by the stellar photospheres,
i.e. below 8$\mu$m. At longer wavelengths some differences appear,  likely
caused by small differences in the average properties of circumstellar envelopes
of mass loosing AGB stars \citep{Bressan06}.
{\bf This uniformity is strengthened by the recent comparison
of the surface brightness distribution in the
Ks and IRS Peak-up 16$\mu$m bands,
of bright ETGs in  Virgo and Coma clusters \citep{Clemens09}}.
NGC~4636 shows also narrow emission
lines but its stellar continuum is that of a passive elliptical.

In contrast,  M~87 shows  a clear MIR excess
which reaches about one order of magnitude at $\approx$20$\mu$m
(see also Perlman et al. 2007). An estimate of the quality of
our IRS extraction process is provided by the independent measure of the
24$\mu$m MIPS flux, plotted in the same figure.


 In summary, inspection of Figure \ref{fig3} shows that M87 stands out in the MIR
spectral window.
In this region, we hypothesize that the stellar populations of M~87 and the other passive
ETGs are similar, and we subtract the spectra of the passive ETGs from that of M~87.
\begin{figure}
\center{\resizebox{0.47\textwidth}{!}{
\psfig{figure=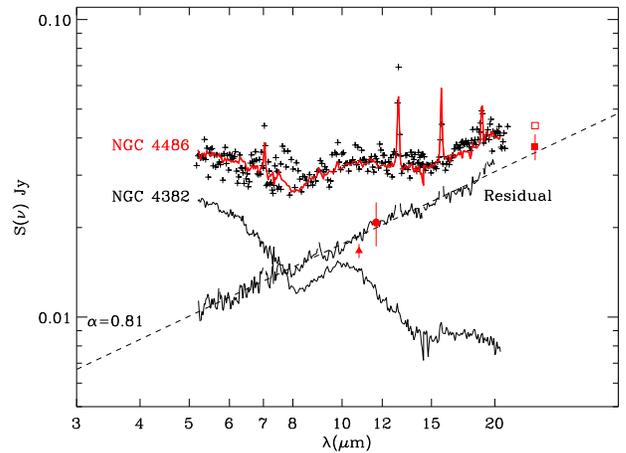,angle=90,clip=}
}}
\caption{MIR residual after subtracting the Ks band normalized template SED of NGC 4382
from that of M~87 (NGC 4486), both indicated in the figure.
The exponent of the power-law fit (dashed line, S$_\nu\propto\nu^{-\alpha}$)
is shown. {\bf The dot and the triangle mark the 11.67$\mu$m and 10.8$\mu$m ground based
nuclear measurements by \citet{Perlman07} and \citet{Perlman01}, respectively}. The  open square is the
M~87 MIPS 24 $\mu$m flux. After correction for the NGC~4382
extrapolated emission, the M~87 MIPS 24 $\mu$m flux lowers to the position indicated by the filled square.
Crosses show another IRS spectrum of M~87 that was extracted from
the archive data (ID 82, PI G.H. Rieke) used by \citet{Perlman07}.}
\label{fig4}
\end{figure}
\begin{deluxetable}{lcccccc }
\centering
\tablecaption{M~87 core MIR power-law (S$_\nu\propto\nu^{-\alpha}$ ) }
\label{tab2}
\tabletypesize{\scriptsize}
\startdata
\hline
Galaxy&$\alpha$&$\alpha_{SL}$&S/F$_{Ks}$&S$_{3.6}$&S$_{11.7}$ \\
Template     &    &     & mJy & mJy  \\
\hline
NGC 4365&0.766$\pm$0.004&0.756$\pm$0.007&0.06&8.9&20.0\\
NGC 4382&0.807$\pm$0.005&0.821$\pm$0.007&0.05&7.7&21.3\\
NGC 4636&0.817$\pm$0.005&0.774$\pm$0.008&0.06&8.1&22.0\\
\enddata
\end{deluxetable}

\section{The residual component}
\label{residual}

To highlight the nature of the MIR excess we  simply subtract the  SEDs of the
passive ETGs from that of M~87. Figure~\ref{fig4} shows the results obtained using the SED of
NGC~4382, but similar results are obtained adopting the other
galaxies as templates for the stellar emission.
The residual, without emission lines that have been masked out,
is well-fitted with a power-law and does not show any evidence of
other particular features reminiscent of additional emission components.
This is even more surprising considering the prominent broad features
visible in the stellar template and given the relative contribution
of the stellar and non-stellar components.
Table~2 summarizes the power-law exponents
(S$_\nu\propto\nu^{-\alpha}$), the fractional contribution of the
power-law in the Ks band, and its absolute flux at 3.6 and 11.7$\mu$m,
for the three templates adopted.
Errors in the exponents are only formal errors of the fits and do not include
errors from the normalizations involved in our procedure.
A realistic uncertainty is provided by the range of slopes
obtained with different galaxy templates.

{\bf Our residual flux (Table~2) is consistent with the high resolution, ground based
nuclear values by \cite{Perlman07} (S$_\nu$= 20.8$\pm $3.5 mJy at 11.67$\mu$m, FWHM=$0\farcs55$, Figure~\ref{fig4} red dot)
and  \cite{Perlman01} (S$_\nu=16.7\pm0.9$ mJy at 10.8$\mu$m, FWHM$=0\farcs46$,  Figure~\ref{fig4} red triangle).}
In the same Figure, we have also plotted the observed MIPS 24~$\mu$m flux, before (open square)
and after (filled square) a 15\% correction due to the contribution of stars, estimated
from the extrapolation of the passive elliptical template.
The MIPS 24~$\mu$m flux is also a time independent measure and falls on the same residual power-law

 The power-law has a slope $\alpha$ between 0.77-0.82, depending on the
adopted stellar population template. This
 value is significantly larger than that found by \citet{Perlman07} ($\sim$0.41).
They fitted  different IRS observations (PID82, Rieke), within
a lower extraction width and using the spectral range 7.5 to 15$\mu$m,
specifically selected to avoid the shorter wavelength region where the contamination
from the stars could be significant.
This large discrepancy could be real, for example due to
a variability of the nuclear component, or could simply be due to
the different method adopted for the analysis.
To understand its origin
we have re-analysed the IRS data set  used by \citet{Perlman07},
with our extraction procedure.
The resulting spectrum is plotted
in Figure \ref{fig4} (crosses).
We notice that our flux values are larger than those published by \citet{Perlman07},
but this is due to the 2.5 times larger extraction width that we have adopted.
Since, apart from this, the two IRS datasets are very similar,
we argue that the different values of the slope arise
from the different approach adopted.
A crude fit to our M~87 SED, in the 7.5 to 15$\mu$m region, gives a slope $\alpha$~$\sim$0.17.
But in our case the strength of the underlying stellar population
is not negligible and changes from 70\% at 5 $\mu$m to 25\% at 15 $\mu$m.
After the subtraction, the slope of the residual is  $\alpha$ $\sim$ 0.8.
Within the smaller aperture used by \citet{Perlman07} the contribution of the stellar population
is certainly less severe than in our case, but not negligible, in particular around 7.5 $\mu$m.
The lack of this correction is most likely the cause of the different slope obtained.

Finally, we notice that the slope we derived is more consistent with
the previous \citet{Perlman01} analysis. Our slope is also similar to the
value found for jet knots A/B and C/G by Shi et al. 2007, and to the slope found at radio
wavelengths when the entire, extended radio source is measured.

\section{Conclusions}
\label{conclusions}

Exploiting our high signal-to-noise {\it Spitzer} IRS spectra and broad band data from the literature,
we have built template SEDs of the passive stellar populations in ETGs.
After a passive stellar component is properly subtracted from our  M~87 IRS spectrum,
the residual, almost independent of the adopted template, is a power-law, and
lacks any signature of the 10$\mu$m feature. This suggests that the 10$\mu$m  bump observed
in the IRS spectrum of M~87 can  be entirely explained in terms of the underlying stellar population,
typical of all other passively evolving Virgo ellipticals (Bressan et al. 2006).
The residual
intensity matches fairly well the high spatial resolution ground based {\it SUBARU}
nuclear measurements at 10.8$\mu$m  and  11.67$\mu$m,
by \citet{Perlman01} and \citet{Perlman07}, respectively.
Since the contribution of stars
at these wavelengths is not negligible, this indicates that our procedure,
even if applied to a large aperture (r$\sim$4\arcsec\ corresponding to $\approx$300 pc at
16 Mpc) is accurate enough to reveal the emission
from the $<$1\arcsec\ core.

The slope of the power law is strikingly similar to that
found for the nearby jet knots and for the large aperture radio data.

The above findings confirm, in a new independent way, that
the inner ($<$1") core of M~87 is dominated
by synchrotron emission, leaving little
room for a torus component \citep{Whysong04,Perlman07}.

Since the signatures of MIR silicate emission
have been detected in  many AGN observed with {\it Spitzer} \citep{Siebenmorgen05,
Shi06}, we suggest that, in some circumstances,
 the underlying stellar population may be an important component to be taken into
consideration.

\begin{acknowledgements}
This work is based on observations made with the {\it Spitzer} Space Telescope, which
is operated by the Jet Propulsion Laboratory, California Institute of Technology
under a contract with NASA. We thank the referee for his comments that
helped to improve the final version of this paper.
L.S., A. B. and G.L. G. thank INAOE for warm hospitality.
A.M. acknowledges support from The Italian Scientists and Scholars of North America Foundation.
We acknowledge contracts ASI-INAF I/016/07/0 and CONACyT 54511.
\end{acknowledgements}

\end{document}